\documentclass[aps,prd,twocolumn,showpacs,superscriptaddress,groupedaddress,nofootinbib]{revtex4}  
\usepackage{graphicx}  
\usepackage{dcolumn}   

\usepackage{graphicx}
\usepackage{amsmath}
\usepackage{amsfonts}
\usepackage{epsfig}
\usepackage{slashed}
\usepackage{braket}
\usepackage{xcolor}
\usepackage{comment}
\usepackage{cases}

\def\bec{\begin{center}}
\def\eec{\end{center}}
\def\beq{\begin{equation}}
\def\eeq{\end{equation}}
\def\bea{\begin{eqnarray}}
\def\eea{\end{eqnarray}}
\def\KD{K\"{a}hler-Dirac }

\def\Psib{\overline{\Psi}}

\def\KD{K\"{a}hler-Dirac }

\def\Psib{\overline{\Psi}}

\def\chib{\overline{\chi}}
\def\Psib{\overline{\Psi}}

\begin{document}
\title{Gauging staggered fermion shift symmetries}

\author{Simon Catterall}
\affiliation{Department of Physics, Syracuse University, Syracuse, NY 13244, USA }
\author{Arnab Pradhan}
\affiliation{Department of Physics, Syracuse University, Syracuse, NY 13244, USA }

\date{\today}

\begin{abstract}
Staggered fermion shift symmetries correspond to translations of the fermion
field within the unit cell of a hypercubic lattice. They satisfy an algebra and in four Euclidean dimensions
can be related to a discrete subgroup of an $SU(4)$ flavor symmetry which plays a crucial role
in showing that staggered fermions lead to a theory of four degenerate Dirac fermions in the continuum limit.  They are associated with
the appearance of certain $Z_2$ valued global parameters.  We propose
a strategy to try to partially gauge these translation
symmetries by allowing these parameters to
vary locally in the lattice. 
To maintain invariance of the action requires the addition
of $Z_2$ valued higher form lattice gauge fields. An analogous procedure can be carried
out for reduced staggered fermions where the shifts correspond to a discrete subgroup
of an $SO(4)$ flavor symmetry.
\end{abstract}

\maketitle

\section{Introduction}
In this paper we will be concerned with the properties of staggered lattice fermions and
in particular the invariance of the staggered action under certain discrete lattice
translations or shift symmetries. It has been known for a long time that these shift symmetries are intimately connected
to the global flavor and axial symmetries of the continuum fermions that arise from the staggered
fields as the lattice spacing is sent to zero \cite{Golterman:1984cy,Golterman:1992yha,Golterman:1984dn,gliozzi1982spinor,KLUBERGSTERN1983447,vandenDoel:1983mf}. One of the easiest ways to understand these features is to recognize that staggered fermions can be viewed as discretizations not of Dirac but
of \KD fermions \cite{Becher:1982ud,Rabin:1981qj,Banks:1982iq,Butt:2021brl,Catterall:2022jky}. 

The \KD equation provides an alternative
to the Dirac equation and is written in terms of the exterior derivative
and its adjoint equation \cite{landau,kahler}. It naturally acts on (Grassmann-valued) antisymmetric tensor fields
(p-forms) and can be formulated on any curved space without the need for a spin
connection or frame. Furthermore, in flat space,
the \KD theory can be written as a theory of matrix valued
fields that satisfy the Dirac equation and whose columns can be interpreted as four
degenerate Dirac spinors.

Massless \KD fermions are particularly interesting as they can be discretized without breaking
a certain classical $U(1)$ symmetry ${\cal G}$ that prohibits mass terms
and without inducing fermion doubling - there is no analog of the Nielsen-Ninomiya theorem in this case \cite{Nielsen:1981hk}. This can be explained by
the fact that $\cal G$ is not the usual chiral symmetry but a twisted cousin that has both a left and right action on the matrix fermion.~\footnote{If one views staggered
fermions as arising from spin diagonalizing the naive lattice Dirac action one usually says that fermion doubling has been reduced from sixteen to four. From the \KD perspective the original continuum theory
already describes four fermions and discretization does not change this -- there is no fermion
doubling from this perspective.}

Furthermore, 
the \KD theory suffers from a mixed $U(1)$-gravitational anomaly that breaks ${\cal G}$ to $Z_4$ once the background space is compactified to a sphere and, remarkably, this anomaly also survives discretization \cite{Catterall:2018lkj,Catterall:2018dns,Catterall:2023nww}.
Attempts to gauge this residual $Z_4$ symmetry in the lattice theory reveals the
presence of a mod 2 't Hooft anomaly. Cancellation of this 't Hooft anomaly requires multiples of two \KD or staggered fields and is a necessary condition to obtain
the massive symmetric phases that have been observed for
staggered fermions \cite{Ayyar:2014eua,Ayyar:2015lrd,Ayyar:2016lxq,Catterall:2015zua,Catterall:2018dns,Catterall:2023nww}. 

In the flat space continuum limit these anomaly free theories
comprise multiples of eight Dirac fermions. If these
spinors live in real or pseudoreal representations of any underlying symmetry group, this is equivalent to multiples of sixteen Majorana fermions - precisely the number that are needed to gap
boundary states in topological superconductors  \cite{Razamat:2020kyf,Fidkowski:2009dba,Morimoto:2015lua,You:2014vea,Wang:2018cai}.

The generator of the ${\cal G}$ symmetry can also be used to define a projector that acts on a \KD field to remove half the degrees of freedom. 
In the continuum flat space theory
it is straightforward to see that this so-called {\it reduced} \KD field
is in fact a chiral theory comprising a doublet of left- and right-handed Weyl
fields transforming under two independent $SU(2)$ global flavor symmetries \cite{Catterall:2022jky,Catterall:2023nww}. In the case of
staggered fermions, an analogous projection yields
reduced staggered
fermions. Thus, it is natural to investigate
the latter lattice theories as a possible way to construct a certain class of chiral lattice
gauge theory \cite{Catterall:2020fep,Catterall:2023nww}.

To elucidate the structure and emergent symmetries of staggered fermion theories, one
must first understand the exact symmetries of the lattice theory. These include rotations, reflections, $\cal G$,
and a set of shift symmetries. The
shifts correspond to lattice translation within the unit hypercube
with appropriately chosen site-dependent phases. These shift
symmetries constrain the form of bilinear mass terms that can arise as a result of quantum corrections. In fact, they function as a discrete subgroup of the continuum axial-flavor symmetries and, in the case of asymptotically free theories
such as QCD, give rise to the usual global $SU(4)_V\times SU(4)_A$ symmetry of four continuum
(massless) Dirac fermions in the continuum limit
\cite{vandenDoel:1983mf,Golterman:1984cy,Golterman:1984dn}.  

We know that $SU(4)_V$ is a non-anomalous symmetry of
continuum theory and hence can be gauged. It is natural to ask whether it is possible
to gauge the corresponding lattice shift symmetries. In this paper, we will
focus primarily on the question of whether there are any
obstructions to coupling the lattice theory to a background
gauge field associated with local shift transformations. In other words, we address
the question as to whether there are 
lattice 't Hooft anomalies associated to these shift symmetries. Our work is therefore closely
related to efforts to extend the ideas of Hooft anomalies to discrete groups and generalized symmetries \cite{Gaiotto:2014kfa}. The question of whether 't Hooft anomalies exist for
translation symmetries of Majorana chains
has also been explored recently in \cite{Seiberg:2023cdc}.

For an {\it onsite} lattice symmetry, which acts on the fields
at a single lattice site, it is trivial to gauge the action by simply inserting the appropriate gauge links between lattice fermions at different sites. If one can do this, then typically
the symmetry has vanishing anomaly. However, one should be careful - it is sometimes
possible for the lattice measure to have an anomalous variation under the symmetry. This is
precisely what happens for the $Z_4$ symmetry discussed earlier.~\footnote{A lack of invariance of the lattice measure also occurs for overlap/Ginsburg-Wilson fermions but in this case the symmetry is not onsite since the generator of the symmetry depends on a covariant difference operator.}

However, in the case of the shift symmetries, 
it is far from obvious how to gauge the lattice action since
the symmetries are {\it offsite} - they
involve lattice translation. 
One cannot simply insert gauge links between sites. Instead, in this paper,
we will try to give a prescription to partially gauge these shift
symmetries using $Z_2$ valued higher-form lattice gauge fields. 

It is important to recognize that the presence of 
't Hooft anomalies can be detected by simply
examining the response of the theory to gauge transformations 
in the presence of a fixed (non-dynamical) background gauge
field. Thus, detecting 't Hooft anomalies in a lattice theory does not
require an understanding of the full phase structure of a theory in the presence of
dynamical gauge fields or even whether that theory has a continuum limit. 
We can, of course, imagine adding kinetic terms for the gauge field and try to understand
the phase structure of the resultant lattice gauge theory too. This is clearly very interesting and important, particularly for reduced staggered fermion theories and their possible connection to chiral
gauge theories. We do consider this possibility towards the end of our paper, but it would require a great deal more work to determine whether such theories have an interesting continuum limit with emergent continuum gauge fields. 

We start with a review of staggered fermions and shift symmetries, emphasizing
the importance of the latter for restoration of continuum flavor symmetry in the
continuum limit. We then describe our strategy for gauging these symmetries and how it
leads to the
appearance of higher form lattice gauge fields. We discuss both full staggered
fermions and reduced staggered fermions where there are additional restrictions on the allowed
shifts. We then
briefly discuss the more general problem of coupling the lattice theory
to dynamical gauge fields and end with a discussion of future directions.

\section{Global shift symmetries of staggered fermions}
The massless four dimensional staggered fermion action takes the form
\begin{equation}S=\sum_{x,\mu}\eta_\mu(x)\chib(x)\left[\chi(x+\mu)-\chi(x-\mu)\right]\label{stag}\end{equation}
where $\eta_\mu(x)=\left(-1\right)^{\sum_{i=1}^{\mu-1} x_i}$ are the usual
staggered fermion phases.
Consider performing a lattice translation or shift transformation in the direction
$\lambda$:
\begin{align}
\chi(x)&\to i\alpha_\lambda\xi_\lambda(x)\chi(x+\lambda)\nonumber\\
\chib(x)&\to -i\alpha_\lambda\xi_\lambda(x)\chib(x+\lambda)
\end{align} with
$\xi_\mu(x)=\left(-1\right)^{\sum_{i=\mu+1}^4 x_i}$ and the global $\alpha_\lambda$ parameter taking its values in the 
group $Z_2\in \{1,-1\}$.
The action transforms to
\begin{align}
S^\prime&=\sum_{x,\mu}\eta_\mu(x)\chib(x+\lambda)\xi_\lambda(x)[
\xi_\lambda(x+\mu)\chi(x+\mu+\lambda)\nonumber\\
&\qquad \qquad \qquad \qquad \qquad \quad-\xi_\lambda(x-\mu)\chi(x-\mu+\lambda)]\nonumber\\
&=\sum_{x,\mu}\eta_\mu(x)\chib(x)\left[\chi(x+\mu)-\chi(x-\mu)\right]\left[\eta_\mu(\lambda)\xi_\lambda(\mu)\right]\end{align}
But it is easy to see from the definitions of the phase factors
that $\eta_\mu(\lambda)\xi_\lambda(\mu)=1$ (see the appendix)
and so the action is invariant under this shift \cite{vandenDoel:1983mf,Golterman:1984dn,Golterman:1984cy}.  
Notice that two successive transformations in the same
direction accomplish a simple translation  $\chi(x)\to \chi(x+2\lambda)$ (up to a trivial change of sign of
$\chi$). 

You can compound these shifts to build other discrete symmetries. For example
the following double shift transformation in the $\rho$,$\lambda$ directions (with $\rho\ne \lambda$)
is also a symmetry 
\begin{align}
\chi(x)&\to \alpha_{\lambda\rho}\xi_\lambda(x)\xi_{\rho}(x+\lambda)\chi(x+\lambda+\rho)\quad \nonumber\\
\chib(x)&\to \alpha_{\lambda\rho}\xi_\lambda(x)\xi_{\rho}(x+\lambda)\chib(x+\lambda+\rho)
\end{align}
where $\alpha_{\lambda\rho}\in Z_2$ and $\rho$,$\lambda$ are not summed over. Using the relation (see the appendix)
\begin{equation}   
\xi_\mu(x)\xi_\nu(x+\mu)+\xi_\nu(x)\xi_\mu(x+\nu)=2\delta_{\mu\nu}\label{alg}
\end{equation}
one can 
show that $\alpha_{\rho\lambda}$ behaves like a $Z_2$ valued antisymmetric tensor.
These double shifts also obey an algebra. 
For example consider the combination of two such
double shifts characterized by $(\mu,\nu)$ and $(\nu,\lambda)$ (we consider just global shifts for now)
\begin{align}\label{twodouble}
    \chi(x)&\to \xi_\mu(x)\xi_\nu(x+\mu)\xi_\nu(x+\mu+\nu)\xi_\lambda(x+\mu+2\nu)\nonumber\\
&\ \ \ \ \ \ \qquad\qquad\qquad\quad\qquad\times\chi(x+\mu+2\nu+\lambda)\nonumber\\
    &= \xi_\mu(x)\xi_\lambda(x+\mu)\chi(x+\mu+\lambda+2\nu)
\end{align}
So two such double shifts give rise to a third distinct $(\mu,\lambda)$ double shift 
plus an ordinary translation by $2\nu$~\footnote{This mixing of lattice translations with would be
internal symmetries is
reminiscent of recent work on Majorana chains with non-invertible symmetries - see \cite{Seiberg:2023cdc}}. The relations given in eqn.~\ref{alg} and
eqn.~\ref{twodouble} are reminiscent of the properties of the Clifford algebra of (Euclidean) Dirac matrices
and suggest a connection to continuum symmetries. To understand this better we turn to the
\KD representation of staggered fermions in the next section.

\section{Continuum interpretation of the shifts}

To get a better understanding of the origin of
these shift symmetries it is helpful
to assemble all the staggered fields in a hypercube into a matrix fermion 
\begin{equation}\Psi(x)=\frac{1}{8}\sum_{b}\chi(x+b)\gamma^{x+b}\label{hyp}\end{equation}
In this expression $b_i$ is a 4-component vector with components $b_i=0,1$ and the sum
extends over all the points in
the unit hypercube with corner $x$. The notation $\gamma^x$ denotes $\gamma^x=\gamma_1^{x_1}\gamma_2^{x_2}\gamma_3^{x_3}\gamma_4^{x_4}$.
Clearly the matrix
field $\Psi(x)$ carries sixteen times as many degrees of freedom as the original
staggered field so not all components of $\Psi(x)$ can be thought of as independent variables. However,
following  \cite{Bock:1992yr} the components of low momentum modes $\Psi(p)$ with $-\frac{\pi}{2}<ap_\mu\le \frac{\pi}{2}$
can be considered independent and we will restrict to these components
in our consideration of the global symmetries of the free theory. The appearance of matrix valued fields is similar to
the so-called
spin-taste basis of staggered fermions \cite{gliozzi1982spinor,DUNCAN1982439,KLUBERGSTERN1983447,Kilcup:1986dg}.  In this latter
construction the matrix lives on a lattice with twice the
lattice spacing with a unit cell that contains
the spin and flavor components of
four Dirac fermions - the latter corresponding to the columns of $\Psi$ \cite{Becher:1982ud,Banks:1982iq}.

As discussed in the introduction, continuum \KD fermions also 
possess a representation in terms of matrix fields. 
The matrix form of the \KD action in flat space is given by
\begin{equation}
S=\int d^4x\,{\rm Tr}\left[\Psib \gamma^\mu\partial_\mu \Psi\right]\label{cont}\end{equation}
where the matrix fermion can be written in terms of its p-form components 
$\chi_{\mu_1\ldots \mu_p}$
\begin{equation}
    \Psi(x)=\sum_p\chi_{\mu_1\ldots \mu_p}(x)\gamma^{\mu_1}\cdots \gamma^{\mu_p}\label{pform}
\end{equation}
Comparing eqn.~\ref{pform} with eqn.~\ref{hyp} one can see that the map between staggered fields and continuum \KD fields is given by
\begin{equation}
\chi(x+b)\to \chi_b(x)
\end{equation} with the non-zero elements of the lattice vector $b$ labeling
the continuum p-form. In fact if we replace the continuum integral in eqn.~\ref{cont} by a sum
over lattice sites and the derivative operator
by a symmetric difference operator, substitute eqn.~\ref{hyp} into this discrete
action, and do the matrix traces, one will
obtain the usual staggered fermion action in eqn.~\ref{stag}. The phase $\eta_\mu(x)$ just reflects the net effect of commuting the extra factor of $\gamma^\mu$ from the displaced
fields $\chi(x\pm \mu)$ through a string of gamma matrices until
it multiplies the explicit factor of $\gamma^\mu$ in the kinetic operator. In this way, staggered
fermions can be explicitly realized by discretization of \KD fermions.

The symmetries of the continuum action in eqn.~\ref{cont} 
are manifest and correspond to the transformations
\begin{align}\Psi\to L\Psi F^\dagger\nonumber\\
\Psib\to F\Psib L^\dagger\end{align}
where $L$ are (Euclidean) $SO(4)$ Lorentz transformations acting on the
spinors that form the columns of $\Psi$ and $F$ are $SU(4)$ flavor
transformations. 
\begin{align}
L&=e^{i\sum_B^6\alpha_B \Gamma_B}\quad \Gamma_B=\{i\gamma_{\mu\nu}\equiv i\gamma_\mu\gamma_\nu\}\quad \mu<\nu\nonumber\\
F&=e^{i\sum_A^{15}\beta_A \hat{\Gamma}_A}\quad  \hat{\Gamma}_A=\{\hat{\gamma}_\mu,i\hat{\gamma}_{\mu\nu},i\hat{\gamma}_5\hat{\gamma}_\mu, \hat{\gamma}_5\}
\end{align}

In general, the antisymmetric tensors of the \KD theory emerge only
when we restrict ourselves to the diagonal subgroup of the Lorentz and flavor symmetries
corresponding to $L=F^\prime$ where $F^\prime$ is the $SO(4)$ subgroup
of $SU(4)$. This procedure identifies the Lorentz Dirac matrices $\gamma_\mu$ with the flavor Dirac matrices $\hat{\gamma}_\mu$. 
In practice, when we employ a staggered fermion discretization of the \KD action,
we realize even less symmetry than this, since the Lorentz group is broken to discrete lattice rotations.

In particular,
the only flavor symmetries that survive discretization correspond to transformations
arising from rotations of the form
$e^{i\alpha_A \frac{\pi}{2}\Gamma^A}=i\alpha_A \Gamma^A$ where $\alpha_A\in Z_2$.
For example, taking $\Gamma^A\equiv\gamma^\lambda$ the lattice matrix fermion transforms
under discrete flavor rotations as
\begin{align}
\Psi(x)&\to \Psi(x+\lambda)i\alpha_\lambda\gamma^\lambda\nonumber\\
\Psib(x)&\to -i\alpha_\lambda\gamma^\lambda\Psib(x+\lambda)
\end{align}
It is not hard to see that this discrete
transformation of the matrix fermion gives rise to the 
staggered shift symmetry described earlier. Using eqn.~\ref{hyp} one finds
\begin{align}
\Psi^\prime &=\frac{1}{8}\sum_b \chi(x+b+\lambda)\gamma^{x+b+\lambda}i\alpha_\lambda\gamma^\lambda\nonumber\nonumber\\
&=\frac{1}{8}\sum_b i\alpha_\lambda\xi_\lambda(x+b)\chi(x+b+\lambda)\gamma^{x+b}
\end{align}
where the phase $\xi_\lambda(x)$ arises because of commuting $\gamma^\lambda$ 
from the right through to its appropriate spot in the string of $\gamma$ matrices.
Clearly, the net effect is to shift $\chi(x)\to i\alpha_\lambda\xi_\lambda(x)\chi(x+\lambda)$ - a shift transformation on the
staggered fermion.
Similarly the double shifts discussed earlier correspond to 
continuum transformations of the
form $e^{\frac{\pi}{2}\alpha_{\lambda\rho}\gamma^\lambda\gamma^\rho}=\alpha_{\lambda\rho}\gamma^\lambda\gamma^\rho$.

In four dimensions there are $32$ discrete shift
symmetries of the staggered action that can be built using these transformations and their inverses. These correspond to the discrete subgroup $\Gamma_4$ of $SU(4)$ given by
\begin{equation}
\Gamma_4=\{\pm I,\pm \gamma^\lambda,\pm i\gamma^\lambda\gamma^\rho,
\pm i\gamma^5\gamma^\lambda, \pm \gamma^5\}
\end{equation}

The non-Abelian character of this group is replicated by the properties of
the staggered phases $\xi_\mu(x)$. As shown in eqn.~\ref{alg} these phases obey an algebra
similar to that of the (Euclidean) gamma matrices and, when these phases are
compounded, they mimic the properties of $\Gamma_4$. We have already seen this in the case of compounding two
double shifts 
and it is easy to see that a similar property 
holds for a combination of double and single shifts (see appendix)
\begin{align}
\label{algeb2}
\xi_\mu&(x+\nu+\lambda)\xi_\nu(x+\lambda)\xi_\lambda(x)\nonumber\\ &\quad\quad\quad\quad\quad-\xi_\lambda(x+\mu+\nu)\xi_\mu(x+\nu)\xi_\nu(x)\nonumber\\
&\quad\quad\quad\quad\quad\qquad=2\left(\delta_{\nu\lambda}\xi_\mu(x)-\delta_{\mu\lambda}\xi_\nu(x)\right)
\end{align}
which corresponds to a piece of the $\Gamma_4$ algebra 
\begin{equation}[\gamma^\mu\gamma^\nu,\gamma^\lambda]=2(\delta_{\nu\lambda}\gamma^\mu-\delta_{\mu\lambda}\gamma^\nu)\end{equation}
Notice again that compounding lattice shifts yields another shift in $\Gamma_4$ up to
a simple translation by a multiple of two lattice spacings -- the lattice shift symmetries generate a 
product of $\Gamma_4$ and the translation group. 

Thus, the lattice shift symmetries realize the action of
the symmetry group $\Gamma_4$ up to translations. But $\Gamma_4$ is a discrete
subgroup of $SU(4)$. Invariance under $\Gamma_4$ and the other exact lattice
symmetries then guarantees restoration of
$SU(4)$ flavor symmetry in the continuum limit -- one cannot write down any relevant
counterterms which are invariant under this discrete subgroup but which break $SU(4)$
\cite{Golterman:1984cy}.~\footnote{Strictly this is true only if we are able to use power counting to assess the relevance of operators. This is fine for asymptotically free
theories like QCD but one should be cautious for theories with fixed points at strong
coupling.}

Actually, there are two
independent sets of shift symmetries of the massless theory. Our discussion so far has
focused on the transformation
\begin{align}
    \chi(x)&\stackrel{\Gamma^V_4}{\to} +i\alpha_\lambda\xi_\lambda(x)\chi(x+\lambda)\nonumber\\
    \chib(x)&\to -i\alpha_\lambda\xi_\lambda(x)\chib(x+\lambda)
\end{align}
corresponding to the
continuum transformation of $\Psi\to \Psi e^{i\frac{\pi}{2}\alpha_\lambda\gamma^\lambda}$ 
and $\Psib\to e^{-i\frac{\pi}{2}\alpha_\lambda\gamma^\lambda}\Psib$. 
The other set of transformations are given by
\begin{align}
    \chi(x)&\stackrel{\Gamma^A_4}{\to} +i\epsilon(x)\alpha_\lambda\xi_\lambda(x)\chi(x+\lambda)\nonumber\\
    \chib(x)&\to +i\epsilon(x)\alpha_\lambda\xi_\lambda(x)\chib(x+\lambda)
\end{align}
corresponding to $\Psi\to \Psi e^{i\alpha_\lambda\frac{\pi}{2}\gamma^5\otimes\gamma^5\gamma^\lambda}$
and $\Psib \to  e^{i\alpha_\lambda\frac{\pi}{2}\gamma^5\otimes \gamma^5\gamma^\lambda}\Psib$. 
These are the analog of ``axial'' transformations - where $\gamma^5$ is replaced by $\gamma^5\otimes \gamma^5$ which has both a right and left action on the matrix fermion. It is the symmetry $\cal G$ discussed in the introduction and becomes
the site parity operator $\epsilon(x)$ of staggered fermions.

\section{Gauging shifts and higher form gauge fields}
\subsection{Single shifts}
To gauge these symmetries, we will let the parameters $\alpha$ depend on
position. We will consider only massive theories where we add the local
mass term $m\sum_x \chib(x)\chi(x)$ to the action. The addition of this term will
break the global shifts generated by $\Gamma^A_4$ and therefore we will not be concerned
with gauging them.

To start we focus on the single shift symmetry and allow $\alpha_\lambda=\alpha_\lambda(x)$  The transformation is now
\begin{align}
    \chi(x)&\to i\alpha_\lambda(x+\lambda)\xi_\lambda(x)\chi(x+\lambda)\label{shift}\nonumber\\
    \chib(x)&\to -i\alpha_\lambda(x+\lambda)\xi_\lambda(x)\chib(x+\lambda)
\end{align}
for a fixed $\lambda$ (it is natural to include a shift of $\lambda$ in the argument of the shift $\alpha_\lambda$).
What happens when a shift of two lattice spacings is made in the same direction ? In this
case 
\begin{equation}\chi(x)\to -\alpha_\lambda(x+\lambda)\alpha_\lambda(x+2\lambda)\chi(x+2\lambda)\label{trans}\end{equation}
Since one wants such a shift to be a simple translation this seems to require
$\alpha_\lambda(x+\lambda)\alpha_\lambda(x+2\lambda)=-1$. However, 
this is a strong constraint since it determines all $\alpha_\lambda(x)$ in direction $\lambda$ from their value on a single slice.
However, there is another possibility -- namely that the factor $\alpha_\lambda(x+\lambda)\alpha_\lambda(x+2\lambda)$ can be interpreted as 
a local $Z_2$ gauge transformation acting on $\chi(x)$. In this way the two lattice spacing shift would yield a simple
translation up to an unobservable $Z_2$ phase. Of course, this requires the introduction of a $Z_2$ link gauge
field $L_\mu(x)$ to keep the action invariant. The gauged action is then given by
\begin{equation}
    S=\sum_{x,\mu}\eta_\mu(x)\chib(x)\left[L_\mu(x)\chi(x+\mu)-L_\mu(x-\mu)\chi(x-\mu)\right]
\end{equation}
In the case of a single local shift $x\to x+\lambda$ with $\alpha_\lambda(x)$ the transformed action is
\begin{align}
S^\prime&=\sum_{x,\mu}\eta_\mu(x)\chib(x)[L_\mu(x)\alpha_\lambda(x)\alpha_\lambda(x+\mu)\chi(x+\mu)\nonumber\\
&\quad \quad \quad\quad -L_\mu(x-\mu)\alpha_\lambda(x-\mu)\alpha_\lambda(x)\chi(x-\mu)]\label{transformed}
\end{align}
Consider first the case $\mu\ne\lambda$. To ensure invariance of such a term requires the addition of new degrees of freedom. 
Specifically, we can 
introduce a plaquette (or 2-form) $Z_2$ gauge field $P_{\mu\lambda}(x)$. Gauge transformations of this object correspond to the appearance of local $Z_2$ phases 
for each boundary link of $P_{\mu\lambda}(x)$. 
Using this freedom we perform the gauge transformation
\begin{equation}P_{\mu\lambda}(x)\to \alpha_\lambda(x)P_{\mu\lambda}(x)\alpha_\lambda(x+\mu)\end{equation}
at the same time as the shift which can then compensate for the factors of $\alpha_\lambda(x)\alpha_\lambda(x+\mu)$ appearing in the first term on the right hand side of eqn.~\ref{transformed}. A similar gauge transformation
of $P_{\mu\lambda}(x-\mu)$ can compensate for the factor $\alpha_\lambda(x-\mu)\alpha_\lambda(x)$ arising in the second term. For this cancellation to work 
the staggered action must be generalized so that each 
fermion link term is dressed by the product of all plaquettes that include that link.
\begin{equation}
    \begin{split}
        S=\sum_{x,\mu}\eta_\mu(x)\,\chib(x)
    \bigg[\bigg(\prod_{\lambda\ne \mu}P_{\mu\lambda}(x)\bigg)L_\mu(x)\chi(x+\mu)\\
    -\bigg(\prod_{\lambda\ne \mu}P_{\mu\lambda}(x-\mu)\bigg)L_\mu(x-\mu)\chi(x-\mu)\bigg]
\end{split}\label{gauged}
\end{equation}
Now we turn to the case $\mu=\lambda$. In this case the action picks 
up factors like $\alpha_\mu(x)\alpha_\mu(x\pm \mu)$.
This cannot be removed by the transformation of a plaquette (2-form)
gauge field $P_{\mu\lambda}(x)$ but can be removed, as described
earlier, by a gauge transformation on the link field $L_\mu(x)$.

To summarize the gauged action given by eqn.~\ref{gauged} is invariant under 
the local $x\to x+\lambda$ shift if the fields transform as 
\begin{align}
    \chi(x)&\to i\alpha_\lambda(x+\lambda)\xi_\lambda(x)\chi(x+\lambda)\nonumber\\
    \chib(x)&\to -i \alpha_\lambda(x+\lambda)\xi_\lambda(x)\chib(x+\lambda)\nonumber\\
    P_{\mu\nu}(x)&\to (1-\delta_{\nu\lambda}-\delta_{\mu\lambda})P_{\mu\nu}(x+\lambda)\nonumber\\
    &\quad + \alpha_\lambda(x+\lambda)P_{\mu\nu}(x+\lambda)\alpha_\lambda(x+\mu+\lambda)\delta_{\nu\lambda}\nonumber\\
    &\quad + \alpha_\lambda(x+\lambda)P_{\mu\nu}(x+\lambda)\alpha_\lambda(x+\nu+\lambda)\delta_{\mu\lambda}\nonumber\\
    L_\mu(x)&\to \left(1-\delta_{\mu\lambda}\right)L_\mu(x+\lambda)\nonumber\\&\quad +\alpha_\lambda(x+\lambda)L_\mu(x+\lambda)\alpha_\lambda(x+\mu+\lambda)
    \delta_{\mu\lambda}
\end{align} 

\subsection{Higher shifts}
If you want to gauge the double shifts 
one employs the transformations
\begin{align}
\chi(x)&\to \alpha_{\rho\lambda}(x+\rho+\lambda)\xi_\rho(x)\xi_\lambda(x+\rho)\chi(x+\rho+\lambda)\nonumber\\
\chib(x)&\to \alpha_{\rho\lambda}(x+\rho+\lambda)\xi_\rho(x)\xi_\lambda(x+\rho)\chib(x+\rho+\lambda)
\end{align}
where $\alpha_{\rho\lambda}(x)$ are $Z_2$ valued parameters (and $\rho\ne\lambda$). 
In this case each term in the fermion action changes by a factor of $\alpha_{\rho\lambda}(x)\alpha_{\rho\lambda}(x\pm\mu)$. 
For each hopping term in direction $\mu$ arising in the fermion action there are two cases to consider corresponding to the situation when neither of the two indices $\rho$ and $\lambda$ equals $\mu$ or 
the case where either $\rho$ or $\lambda$ equals $\mu$. 

Let's first consider the case when
$\rho\ne\mu$ and $\lambda\ne\mu$. In this case 
one can cancel off the factor arising in the
transformed fermion kinetic term 
using a gauge transformation of a cubical (3-form) $Z_2$ gauge field
$C_{\mu\rho\lambda}(x)$ that transforms by local
$Z_2$ factors associated with each $\rho\lambda$-plaquette face. 
\begin{equation}
    C_{\mu\rho\lambda}(x)\to \alpha_{\rho\lambda}(x)C_{\mu\rho\lambda}(x)\alpha_{\rho\lambda}(x+\mu)
\end{equation}
In four dimensions there are precisely three such
cubes per link. 

In the second case, one of the directions $\rho$ or $\lambda$ is equal to
the hopping term index $\mu$. Taking the example $\mu=\rho$ the kinetic term acquires the factor
$\alpha_{\mu\lambda}(x)\alpha_{\mu\lambda}(x\pm\mu)$. There are again three of these terms
per link $\mu$. And to retain invariance under such a double shift requires the plaquette (2-form) field associated to this link to transform 
\begin{equation}
    P_{\mu\lambda}(x)\to \alpha_{\mu\lambda}(x)P_{\mu\lambda}(x)\alpha_{\mu\lambda}(x+\mu)
\end{equation} 

The final fermion kinetic term invariant under both single and double
shifts now reads

\begin{align}
S=\sum_{x,\mu}&\eta_\mu(x)\,\chib(x)\nonumber\\
&\times\bigg[\bigg(\prod_{\lambda\ne\rho\ne \mu}C_{\mu\rho\lambda}(x)\bigg)\bigg(\prod_{\lambda\ne\mu} P_{\mu\lambda}(x)\bigg)L_\mu(x)\chi(x+\mu)\nonumber\\
    &\quad-\bigg(\prod_{\lambda\ne\rho\ne \mu}C_{\mu\rho\lambda}(x-\mu)\bigg)\bigg(\prod_{\lambda\ne\mu} P_{\mu\lambda}(x-\mu)\bigg)\nonumber\\
    &\qquad\times L_\mu(x-\mu)\chi(x-\mu)\bigg]
    \label{gauge2}
    \end{align}

Thus the lattice action can
be made invariant under a local double shift $x\to x+\rho+\lambda$ if the fields
transform as follows:
    \begin{align}
    \chi(x)&\to \alpha_{\rho\lambda}(x+\rho+\lambda)\xi_\rho(x)\xi_\lambda(x+\rho)\chi(x+\rho+\lambda)\nonumber\\
    \chib(x)&\to \alpha_{\rho\lambda}(x+\rho+\lambda)\xi_\rho(x)\xi_\lambda(x+\rho)\chi(x+\rho+\lambda)\nonumber\\
    C_{\mu\nu\delta}(x)&\to (1-\delta_{\rho\mu}-\delta_{\lambda\mu})\alpha_{\rho\lambda}(x+\rho+\lambda)\nonumber\\&\times C_{\mu\nu\delta}(x+\rho+\lambda)\alpha_{\rho\lambda}(x+\rho+
    \lambda+\mu)\nonumber\\
    &\times(\delta_{\nu\rho}\delta_{\delta\lambda}+\delta_{\nu\lambda}\delta_{\delta\rho})\nonumber\\
    P_{\mu\nu}(x)&\to \left(1-\delta_{\nu\rho}-\delta_{\nu\lambda}\right)P_{\mu\nu}(x+\rho+\lambda)\nonumber\\
    &+\alpha_{\rho
    \lambda}(x+\rho+\lambda)P_{\mu\nu}(x+\rho+\lambda)\nonumber\\
    &\times\alpha_{\rho\lambda}(x+\rho+\lambda+\mu)\left(\delta_{\nu\rho}\delta_{\mu\lambda}+\delta_{\nu\lambda}\delta_{\mu\rho}\right)\nonumber\\
    L_\mu(x)&\to L_\mu(x+\rho+\lambda)
\end{align}
In the above expressions, to simplify things, we have taken the index
$\mu$ to correspond to the hopping direction. In addition, in the case of a single shift $x\to x+\lambda$, the $C$ field must transform by
a simple translation 
\begin{equation}  
C_{\mu\nu\delta}(x)\to C_{\mu\nu\delta}(x+\lambda)\end{equation}

Clearly, this process can be continued to higher shift symmetries using corresponding
$Z_2$ valued lattice (p+1)-form gauge fields to gauge $p$-shift symmetries. In four dimensions
one will need a hypercubical lattice 4-form field $H$ to gauge triple shifts (together with the 
cubical 3-form gauge field $C$ to handle the case when one of the shift indices coincides with the hopping term index). Naively one might think that there would be no way to gauge
4-shifts since one cannot write down a 5-form field in a four dimensional lattice. However, such a field is not needed. This can be seen by performing a 4-shift on the action. The action
changes by the local factor
\begin{equation}
    \alpha_{1234}(x)\alpha_{1234}(x+\mu)
\end{equation}
But in this expression the index $\mu$ must equal one of the indices on $\alpha$. We find ourselves, therefore,
in the situation where only the 4-form field $H$ is needed to compensate for this factor.

We should add a caveat at this point. Our strategy for gauging the shift symmetries
is to allow the $Z_2$-valued $\alpha$ parameters to depend on position. However,
we do {\it not}
allow the {\it type} of shift i.e. single, double, etc. to depend on position. However, gauging
a discrete group like $\Gamma_4$ would usually require such a strategy. Thus, our proposal is, in some sense, only a partial gauging of the discrete symmetry.

\section{Gauging shifts for reduced staggered fermions}
Using the site parity operator $\epsilon(x)=\left(-1\right)^{\sum_i x_i}$
we can introduce projectors of the form 
\begin{equation}P_\pm=\frac{1}{2}\left(1\pm \epsilon(x)\right)\end{equation} and show that the  massless
staggered fermion action decomposes into two independent parts. Retaining just one of
these leads to an action for a {\it reduced} staggered fermion of the form
\begin{equation}S=\sum_{x,\mu}\eta_\mu(x)\chib_+(x)\left[\chi_-(x+\mu)-\chi_-(x-\mu)\right]\end{equation}
where $\chi_+=P_+\chi$ etc \cite{vandenDoel:1983mf,Bock:1992yr,SHARATCHANDRA1981205}. 
Equivalently we can relabel $\chib_+$ as just $\chi_+$ and write
\begin{equation}
    S=\frac{1}{2}\sum_{x,\mu}\eta_\mu(x)\chi(x)\left[\chi(x+\mu)-\chi(x-\mu)\right]
\end{equation}
In the continuum matrix theory, the projected fields are $\frac{1}{2}\left(1\pm \gamma^5\Psi\gamma^5\right)$ and are equivalent to reduced \KD fermions \cite{Catterall:2023nww,Bock:1992yr,vandenDoel:1983mf}.
The continuum flavor symmetry is then just 
$SO(4)\times U(1)$ since these are the only generators that commute with $\gamma^5$ in the projector. Thus, the discrete subgroup of flavor symmetries for the reduced field is
\begin{equation}
G=\{\pm I, \pm i\gamma^\mu\gamma^\nu, \pm \gamma^5\}\end{equation} 
This implies that the only shift symmetries that survive are the double and 4-shifts.
In practice we restrict our attention to the former and consider a local double
shift of the form~\footnote{A simple local mass term vanishes for a single
flavor of reduced fermion but it is straightforward to construct a mass or Yukawa term
for multiple flavors of reduced fermion.}
\begin{equation}
    \chi(x)\to \epsilon(x)\alpha_{\lambda\rho}(x+\lambda+\rho)\xi_\lambda(x)\xi_\rho(x+\lambda)\chi(x+\lambda+\rho)\label{so4}\nonumber\\
\end{equation}
These transformations
can be gauged by introducing both a plaquette field $P_{\mu\nu}(x)$
and a cube-based $Z_2$ gauge field $C_{\mu\nu\lambda}(x)$ as discussed earlier.
Notice that the absence of any obstruction to gauging the double shifts in the
lattice theory is consistent with the absence of any continuum 't Hooft anomaly
for the continuum $SO(4)$ symmetry.

\section{Dynamical lattice gauge fields}
Up to this point we have focused on 
gauging the shift symmetries in the presence of a background gauge field. We have shown that
this may be done by employing higher form
lattice gauge fields. 

It is natural to go on to consider the theory
in the presence of dynamical lattice gauge fields. To do this, we first need to add
kinetic terms for the gauge fields.
A Wilson plaquette term can be added for the link field $L_\mu(x)$ as usual
\begin{equation}
    S_L=-\beta_L\sum_{x,\mu<\nu}F_{\mu\nu}
\end{equation}
where 
\begin{equation}
    F_{\mu\nu}(x)=L_\mu(x)L_\nu(x+\mu)L_\mu(x+\nu)L_\nu(x)\end{equation}

Similarly, one can add a gauge invariant
kinetic term for the 2-form gauge field $P_{\mu\lambda}(x)$ by taking the product of the plaquette
fields over all faces of a cube in the lattice and subsequently
taking a sum over all cubes. This generalizes the familiar Wilson plaquette 
operator used for gauge links \cite{Lipstein:2014vca}.
\begin{equation}
    S_P=-\beta_P\sum_{x,\mu<\nu<\lambda}F_{\mu\nu\lambda}(x)
\end{equation}
where $F_{\mu\nu\lambda}(x)$ is a product over all faces of a cube with corner at $x$ and with basis vectors along the $\mu$, $\nu$
and $\lambda$ directions
\begin{align}
F_{\mu\nu\lambda}(x)&=P_{\mu\nu}(x)P_{\mu\lambda}(x)P_{\nu\lambda}(x)P_{\nu\lambda}(x+\mu)\nonumber\\
&\times P_{\mu\lambda}(x+\nu)P_{\mu\nu}(x+\lambda)\end{align}

A kinetic operator for the 3-form
$C_{\mu\rho\lambda}(x)$ would consist of a product over all eight cubical
faces of a hypercube.
\begin{equation}
    S_H=-\beta_C\sum_{x,\mu<\nu<\rho<\lambda} F_{\mu\nu\rho\lambda}(x)
\end{equation}
where
\begin{align}
    F_{\mu\nu\rho\lambda}(x)&=C_{\mu\nu\rho}(x)C_{\mu\nu\lambda}(x)C_{\mu\rho\lambda}(x)C_{\nu\rho\lambda}(x)C_{\mu\nu\rho}(x+\lambda)\nonumber\\
    &\quad\quad\quad\times C_{\mu\nu\lambda}(x+\rho)C_{\mu\rho\lambda}(x+\nu)C_{\nu\rho\lambda}(x+\mu)
\end{align}
Notice that it is not possible to write down a field strength for the 4-form field $H$ in four
dimensions.

To understand the nature of the continuum limit in the presence of these dynamical
fields, one must first map out the phase diagram of the lattice theory in the space of
gauge couplings $(\beta_L,\beta_P,\beta_C)$. Continuum limits can only be defined in the vicinity
of continuous phase transitions and so the first goal would be to determine whether
such points exist and the nature of the corresponding critical exponents which 
determine the scaling needed to take a continuum limit. One can attempt to get analytic understanding on the phase diagram through duality transformations along the lines of \cite{RevModPhys.52.453}. In fact the quenched 4-form sector in four
dimensions is trivial -- the absence of a kinetic term allows one to adopt
a gauge in which the 4-form gauge field is set to unity locally. Additionally,
it is possible to rewrite the 3-form sector
as a theory of non-interacting field strengths which renders it
easy to show that it exhibits no 
phase transitions.  The quenched 2-form $Z_2$ system
can be shown to be dual to the 4d Ising model and hence has a continuous phase transition. 
This leaves only the dynamics of the gauge link $L$ to be determined - in this case the
system is thought to undergo a first order transition. These results,
while applying strictly only to the quenched theory, suggest the task of locating physically
relevant critical points in the full theory, while difficult and beyond the scope of the
current work, may not be insurmountable. 

In the case of global shift symmetries, we know that these symmetries of the staggered
fermion enhance to the usual global $SU(4)$ flavor 
symmetry of four Dirac fermions in the continuum limit. Thus, one might conjecture
that a successful gauging of the shift symmetries might yield
a continuum theory
with emergent $SU(4)$ gauge symmetry. One can express similar hopes that the continuum limit
of a lattice theory of reduced staggered fermions with gauged
shifts might admit an emergent $SO(4)$ gauge
symmetry.

However, if $SU(4)$ or $SO(4)$ symmetry is to emerge from such a lattice
model, the higher form gauge fields must reduce to 0-form symmetries in the continuum
limit. The simplest way for this to happen is for
some of the additional
spacetime indices carried by the lattice gauge fields to be realized  
as internal flavor indices as the lattice spacing is sent to zero.
\begin{align}
    P_{\mu\lambda}(x)&\stackrel{a\to 0}{\to} P_\mu^{\lambda}(x)\nonumber\\
    C_{\mu\rho\lambda}(x)&\stackrel{a\to 0}{\to} C_\mu^{\rho\lambda}(x)
\end{align}
where the superscript indices are now to be interpreted as operating on the internal
flavor space. 

In addition, for the lattice gauge symmetries to enhance to the continuum gauge symmetry
at such a critical point, it is necessary to assume that some form
of universality holds which would
ensure that the domain of the $\alpha$ variables enhances from $Z_2$ to the
real line in the continuum limit. This is certainly expected to be the case in the situation
where the symmetries are global and where the
theory is defined around a Gaussian fixed point. The open question is
whether it continues to hold when these shift symmetries are gauged and where it is
not guaranteed that the
continuum limit is taken in the vicinity of such a free fixed point.

\section{Summary}

In this paper we have described a proposal to (partially) gauge the shift symmetries of
staggered fermions. These shift symmetries correspond to
translations within the unit hypercube of a lattice
and can be related to a discrete
subgroup of the continuum flavor symmetry of four degenerate
Dirac fermions~\footnote{In this paper we focus exclusively on
four (Euclidean) dimensions but clearly the approach we use can
be easily generalized 
to any number of dimensions.}. 

They furnish an explicit example of a lattice theory where elementary lattice translations
give rise to global internal symmetries in the continuum limit
and where lattice 't Hooft anomalies may play an important role in understanding the possible phases of the theory \cite{Seiberg:2023cdc,Cheng:2022sgb}. To understand this
result, it is useful to recognize that staggered fermions can be thought of as
discrete approximations to \KD fermions 
whose {\it twisted} symmetry group corresponds to a diagonal
subgroup of the Lorentz and flavor symmetries. The shift symmetries then correspond to
a discrete subgroup of this twisted rotation group and naturally connect
spacetime symmetries with internal symmetries. 

The gauging procedure necessitates the introduction of $Z_2$-valued
higher-form lattice gauge fields. This construction works for shifts that correspond to
discrete subgroups of the flavor symmetry, requires only the introduction of background
gauge fields, and holds for non-zero lattice spacing.
However, we do not know whether it is possible to extend
these arguments to show that these staggered theories admit emergent continuous gauge
symmetry in the continuum limit.
To do this, one would need to 
show that the phase diagram of the lattice theory with dynamical
$Z_2$ gauge fields possesses an appropriate continuous
phase transition where such a continuum limit can be taken. And proving that 
the discrete gauge symmetry discussed in this paper
truly enhances to a continuous symmetry near such a critical point would be an additional non-trivial problem. This issue is key to the question of whether the reduced theory offers a
route to a chiral gauge
theory in the continuum limit \cite{Catterall:2022jky,Catterall:2023nww}. We plan to extend our
work in these directions in the future.

\section*{Acknowledgements}
This work was supported by the US Department of Energy (DOE), 
Office of Science, Office of High Energy Physics, 
under Award Number DE-SC0009998. SMC would like to acknowledge useful talks with
Srimoyee Sen, Shailesh Chandrasekharan, Aleksey Cherman and Yi-Zhuang You.

\begin{appendix}
\section{Identities}
Using
\begin{equation} \eta_\mu(x)=\left(-1\right)^{\sum_{i=1}^{\mu-1} x_i}
\end{equation}
and
\begin{equation}
    \xi_\lambda(x)=\left(-1\right)^{\sum_{i=\lambda+1}^4 x_i},\label{chidef}
\end{equation}
we see that for $\mu-1\geq\lambda$, $\eta_\mu(\lambda)=\xi_\lambda(\mu)=-1$. And for $\mu-1<\lambda$, $\eta_\mu(\lambda)=\xi_\lambda(\mu)=1$. This gives the identity
\begin{equation}
    \eta_\mu(\lambda)\xi_\lambda(\mu)=1.
\end{equation}
Notice that eqn. \ref{chidef} implies $\xi_\mu(x)=\xi_\mu(x+\mu)$, hence $\xi_\mu(x)\xi_\mu(x+\mu)=1$. For $\mu\neq\nu$, 
\begin{align}
\xi_\mu&(x)\xi_\nu(x+\mu)+\xi_\nu(x)\xi_\mu(x+\nu)\nonumber\\
&= \left(-1\right)^{\sum_{i=\mu+1}^4 x_i}\left(-1\right)^{\sum_{i=\nu+1}^4 (x+\mu)_i} + (\mu \leftrightarrow \nu).
\end{align}
From the above expression it is evident that the two terms differ by a factor of $-1$ for the two possible cases, $\mu\geq\nu+1$ and $\nu\geq\mu+1$. Hence we obtain the identity
\begin{equation}
    \xi_\mu(x)\xi_\nu(x+\mu)+\xi_\nu(x)\xi_\mu(x+\nu)=2\delta_{\mu\nu}.
\end{equation}
Using eqn. \ref{chidef}, we see that the two terms on the lefthand side of eqn. \ref{algeb2} differ by a factor of $\left(-1\right)^{\sum_{i=\mu+1}^4 \lambda} \times \left(-1\right)^{\sum_{i=\nu+1}^4 \lambda} \times \left(-1\right)^{\sum_{i=\lambda+1}^4 \nu+\mu}$. We restrict to the case $\mu\neq\nu$ to avoid a double shift along the same direction. For $\mu\neq\nu\neq\lambda$, this factor is $+1$ for all possible orderings, hence the lefthand side of this
expression vanishes. For $\mu=\lambda$, this factor simplifies to -1 for the two possible cases, $\nu>\lambda$ or $\nu<\lambda$. Let us first consider the case $\nu>\lambda$, then the first term of eqn. \ref{algeb2} simplifies to
\begin{align}
\xi_\mu&(x+\nu+\lambda)\xi_\nu(x+\lambda)\xi_\lambda(x) \nonumber\\
&= \xi_\lambda(x+\nu)\xi_\nu(x)\xi_\lambda(x) = \left(-1\right)^{\sum_{i=\lambda+1}^4 \nu}\xi_{\nu}(x)=-\xi_{\nu}(x)\nonumber
\end{align}
The second term of Eq. \ref{algeb2} then differs from the first by a factor of -1 as we just argued and hence the equality holds. Repeating the same for $\nu<\lambda$ leads to the same result. For $\nu=\lambda$ the same analysis yields $+2\xi_\mu(x)$ on the lefthand side. This proves eqn. \ref{algeb2}.

\end{appendix}

\bibliography{shift}

\end{document}